\documentclass[10pt,letterpaper,twocolumn]{article} 

\usepackage{ol2}
\usepackage[draft,implicit=false]{hyperref}
\usepackage{amsmath}
\usepackage{amssymb}

\begin{document}

\twocolumn[ 

\title{Transition from  non-resonant to resonant random lasers \\ by the  geometrical confinement of  disorder}


\author{ N. Ghofraniha$^{1,*}$, I. Viola$^{2}$, A. Zacheo$^{3}$, V. Arima$^4$, G. Gigli$^5$ and C. Conti$^6$}

\address{
$^1$ Institute for Physical Chemical Processes (IPCF-CNR), UOS Roma Kerberos, Universit\`{a} La Sapienza,\\ P. le A. Moro 2,
I-00185, Rome, Italy\\
$^2$ National Nanotechnology Laboratory, Institute Nanoscience-CNR (NNL, CNR-NANO),  I-73100 Lecce,
 Italy \\and c/o Department of Physics, La Sapienza University, Rome, Italy\\ 
$^3$ Dip. di Matematica e Fisica "Ennio de Giorgi", Universit\`{a} del Salento, I-73100, Lecce, Italy\\
$^4$ National Nanotechnology Laboratory, Institute Nanoscience-CNR (NNL, CNR-NANO), I-73100 Lecce, Italy\\
$^5$ National Nanotechnology Laboratory, Institute Nanoscience-CNR (NNL, CNR-NANO),  Lecce,
Italy \\ and c/o Department of Physics, La Sapienza University, Rome; Universit\`{a} del Salento, Dip. di Matematica e Fisica "Ennio de Giorgi" and Italian Institute of Technology (IIT), Centre
for Biomolecular Nanotechnologies,  Lecce, Italy\\
$^6$ Institute for Complex Systems (ISC-CNR) and Department of Physics, University Sapienza, P.le A. Moro 5, I-00185, Rome, Italy\\

$^*$neda.ghofraniha@roma1.infn.it
}

\begin{abstract}We report on a novel kind of transition in  random lasers induced by the geometrical confinement of the emitting material.  
Different dye doped paper devices with controlled geometry are fabricated by soft-lithography
and show two distinguished behaviors in the stimulated emission: in the  absence of boundary constraints the  energy threshold decreases 
for larger  laser volumes showing the typical trend of diffusive 
{\it non-resonant} random lasers,
 while when the same material in lithographed into channels, the walls act as cavity and  the  {\it resonant} behavior typical of standard lasers is  observed.  
The experimental results are consistent with the general theories of random and standard lasers and a clear phase diagram of the transition is reported. 
\end{abstract}


 ] 

\noindent Inhomogeneity of refractive index in materials provokes the scattering of electromagnetic waves with wavelength $\lambda$ comparable to the characteristic size of the refraction variation.
While ordered distribution of the scatterers gives  rise to the typical  wave interference fringes known as Bragg diffraction, disordered distribution induces the random propagation of light 
in space and a more complex interference pattern. This simple effect
 has promoted in the last decades the development of random photonics with novel fascinating physical phenomena and relevant applications~\cite{Wie13}.
Among all, the severe effects of disorder have been widely investigated in nonlinear 
 optical wave propagation, as shock waves~\cite{Gho12,Gen12}, beam filamentation~\cite{Gho06} and soliton formation~\cite{Fol10}; 
 moreover ultra-focusing of light through disordered media is nowadays paving the way to novel
 imaging techniques~\cite{Vel10} and
 Anderson localization of light in the regime of high scattering strength has been considered~\cite{Wie97,Seg13,Spe13}. 
Another intriguing effect of ``disordered light'' is represented by random lasers (RLs), that are light  sources made by a gain medium whose spontaneous emission 
is amplified by the feedback mechanism due to multiple 
scattering. This is achieved by
 adding powders or nano-particles~\cite{Wiersma08,Van07} or by the intrinsic supramolecular structure of the gain material~\cite{Gho13}.  
Depending on the emission properties two main groups of RLs have been classified: RLs with intensity non-resonant feedback characterized, above the threshold,  by a drastic narrowing of the emission 
spectrum in a broad and featureless peak centered at the highest gain frequency;  RLs with field resonant feedback,  presenting multiple sub-nanometer narrow peaks above the amplification line.
Mainly, the formers have been realized  in weakly scattering materials with $l_t \gg \lambda$, with $l_t$ the transport mean free path, and the latter in highly scattering systems with $ l_t \geq \lambda $. 
Interesting exceptions  have also been  reported signaling the crucial role of the pump size, shape~\cite{Wu08} and direction~\cite{Leonetti2011}
as well as the amplifying material~\cite{Muj04}.
Hence, the classification of RLs is still an open question.
More generally, when $l_t\lesssim \lambda$/2$\pi$, the separated emission peaks are monochromatic and identified
with the  localized and long-lived eigenmodes of the Maxwell's equations in the passive system~\cite{Vanneste01,Seb02,Con08}. However, this situation is experimentally 
hardly obtainable especially in 3-dimensional materials 
and most of the experiments reporting resonant peaks are far from this regime. In all other scattering regimes  the eigenmodes are not standing waves 
with defined phase but are short-lived ``quasimodes'', a subset of which can give life to  spatially extended lasing modes across the pumped system~\cite{Vanneste07}.   
The scattering strength and/or pumping conditions  control the  spatial extension of such subset: in the case of a large number of equally excited 
activated modes, they  merge to a continuous broadband;
while in the situation of a reduced amount of modes, the emission spectra present few distinguishable non-monochromatic resonances. 

In this Letter  we introduce  a novel kind of  resonance in  weakly scattering RLs, that is not due to the partial localization of the extended lasing modes, leading to a fine spectral structure,
but is induced by the geometrical confinement of the RL material in micro-channels. 
Our system is an innovative type of RL made of paper channels impregnated by a dye solution, the micro-fluidic channels are realized by soft lithographic methods, that allow the fine control of their size.
Although we observe single broad laser peaks in all cases, being the sample weakly scattering with $l_t=(36\pm4)\mu$m,
we show that there is a net transition in the behavior of the lasing threshold once the RL is geometrically confined  and we report a clear phase diagram of the laser threshold versus sample volume.
When the paper is without channels the typical RL behavior~\cite{vanSoest99} is observed  while  in presence of the channels  the threshold energy trend shows the characteristics of a 
standard laser with the channel walls acting as resonator. 
By recalling the same definitions introduced by Letokhov in 1968 in the first presentation  of  ``quantum generators''~\cite{Let68}, we name this transition: {\it non-resonant} (random laser) to {\it resonant} (standard laser).

\begin{figure}
\includegraphics[width=4cm]{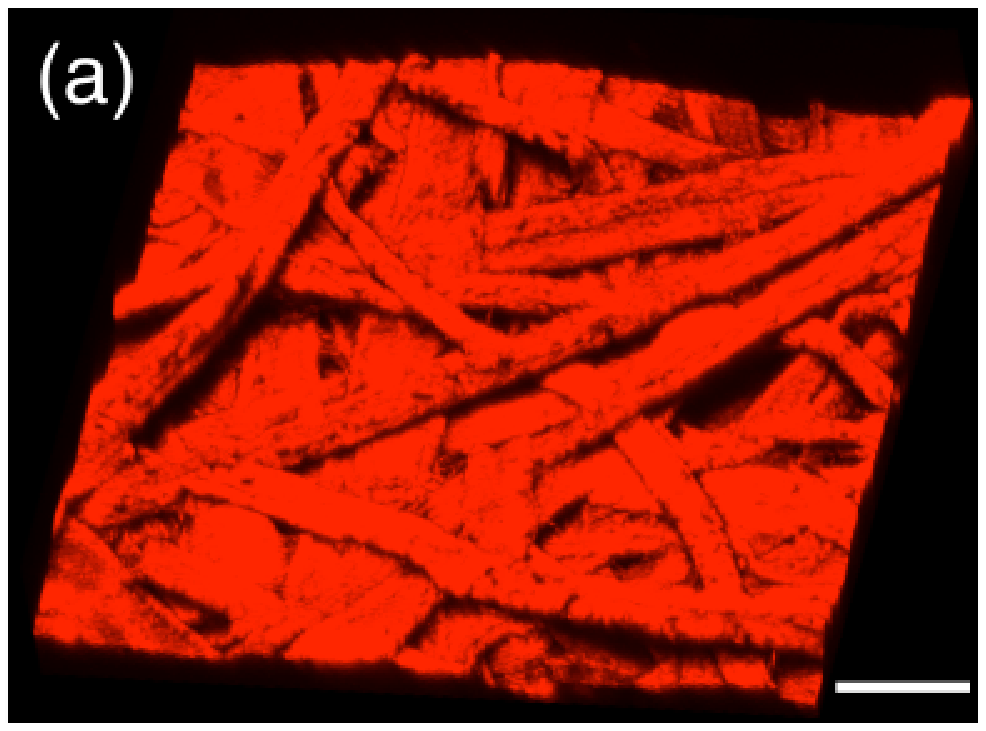}
 \includegraphics[width=4cm]{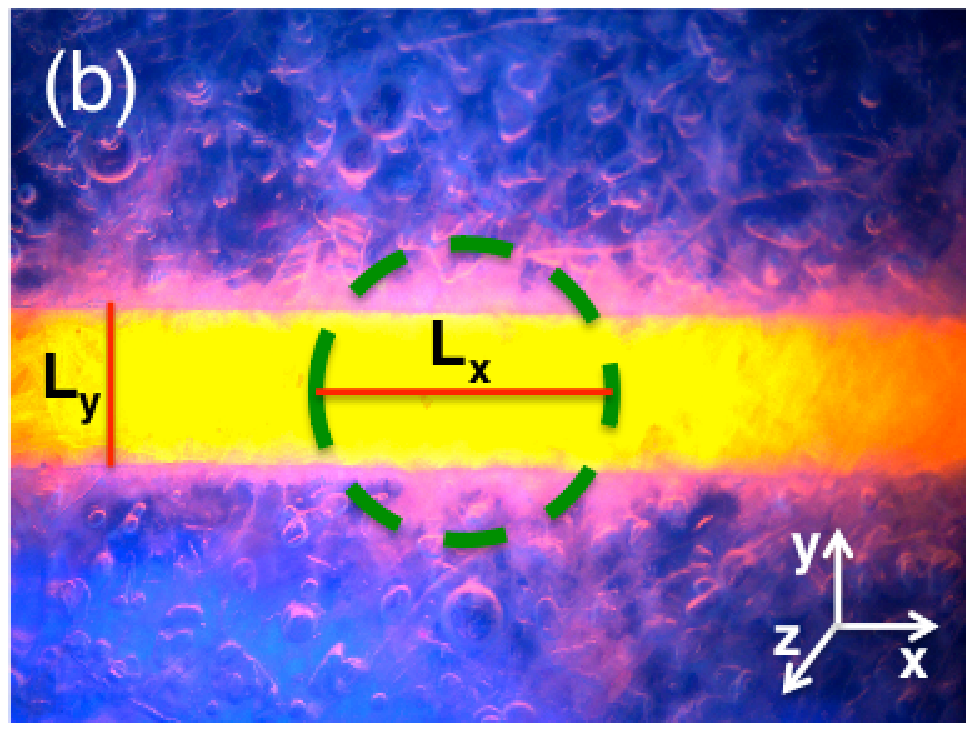}
 \caption{(Color online)  
(a) Confocal microscopy image (3-D reconstruction) showing the paper mash impregnated by  dye solution, the scale bar is 50$\mu$m. 
(b) Fluorescence image of the  micro-fluidic channel with $L_y$=300$\mu$m, the dashed circle indicates the pump spot with diameter $L_x$.} 
\label{fig1}
\end{figure}

Our samples are made of chromatographic paper impregnated by a 1mM solution of Rhodamine B and Ethilene-glycol
and they all have a thickness of  $L_z=100\mu$m. Once in contact, for capillarity, the fluid dye solution diffuses inside the mash of the paper filling all the volume available, giving rise to a  
homogeneous material able to   emit radiation because of the dye presence and to scatter the light because the paper is made of cellulose with a typical index refraction of 1.47 at the visible frequencies.  
We estimate a  light mean transport  path of $l_t=(36\pm4)~\mu$m  at 532~nm wavelength by measuring the transmittance of the paper for various $L_z$ and fitting it with exponential decay.
A representative confocal microscopy image of the microscopic structure is shown in fig.~\ref{fig1}-a. 

Soft lithographic techniques~\cite{Siegel10, Martinez08} are used to fabricate different microfluidic channels with controlled   width $L_y$ of 200~$\mu$m, 300~$\mu$m and  600~$\mu$m, 
an example is illustrated in fig.~\ref{fig1}-b and details of the fabrication are given elsewhere. We use confocal microscopy both to check the homogeneous distribution of the dye
solution, by measuring the fluorescence spectra at several points of the samples along the $x$, $y$, $z$ directions, and to estimate the effective thickness of the dyed paper $L_z' < L_z$ due to the capillarity
effect. This is obtained by measuring the depth of fluorescence from the z-stack confocal images with a precision of 3~$\mu$m.\\
The samples are pumped along the $z$ direction by a frequency doubled Q-switched  Nd:YAG pulsed laser emitting at $\lambda$=532~nm, 
with 10~Hz repetition rate, 6~ns pulse duration and with 8~mm beam diameter. In all the measurements 
 the laser  is focused down to a spot of diameter $L_x=600\mu$m (see fig.~\ref{fig1}-b)
and the transmitted radiation is collected after  pump filtration and focalization into an optic fiber connected 
 to a  spectrograph equipped with electrically cooled CCD array detector. 
 To avoid bleaching effects, we mount our samples on a translator with micro metric control  and take each single pulse  emission spectrum  from a  distinct  point of the channels.
 No pulse to pulse variations   in the emission are observed for fixed pumping.\\
 We estimated the absorption and emission cross sections of the dye solution respectively as $\sigma_a=\alpha / N\simeq$ 1.7 $10^{-16}cm^2$
and $\sigma_e=g/N\simeq$  5  $10^{-17}cm^2$, being the measured absorption     $\alpha=10 mm^{-1}$,  the gain coefficient $g\simeq 30$ cm$^{-1}$ (assuming that a 1\% in volume of Rhodamine B
solution in diethylene-glycol at 0.5 nJ/$\mu m^2$ has a gain of
100 cm$^{-1}$, as measured in [24]) and the dye molecule number density $N\simeq$ 6 $10^{17}$ cm$^{-3}$. Moreover the  paper samples 
are isotropic as evidenced by the investigation of the scattered intensity at different angles.

\begin{figure}[h!]
\includegraphics[width=7cm]{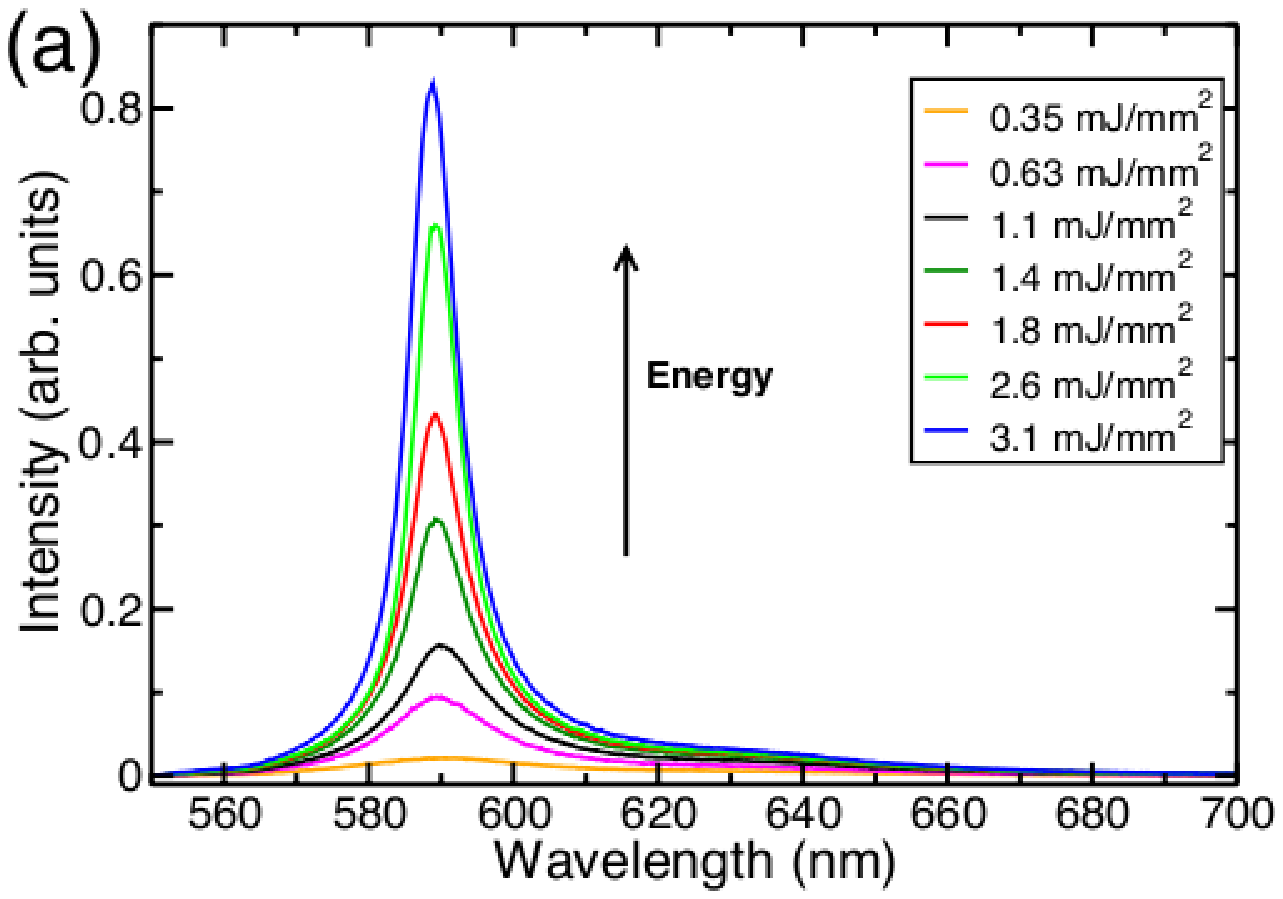}
\includegraphics[width=7cm]{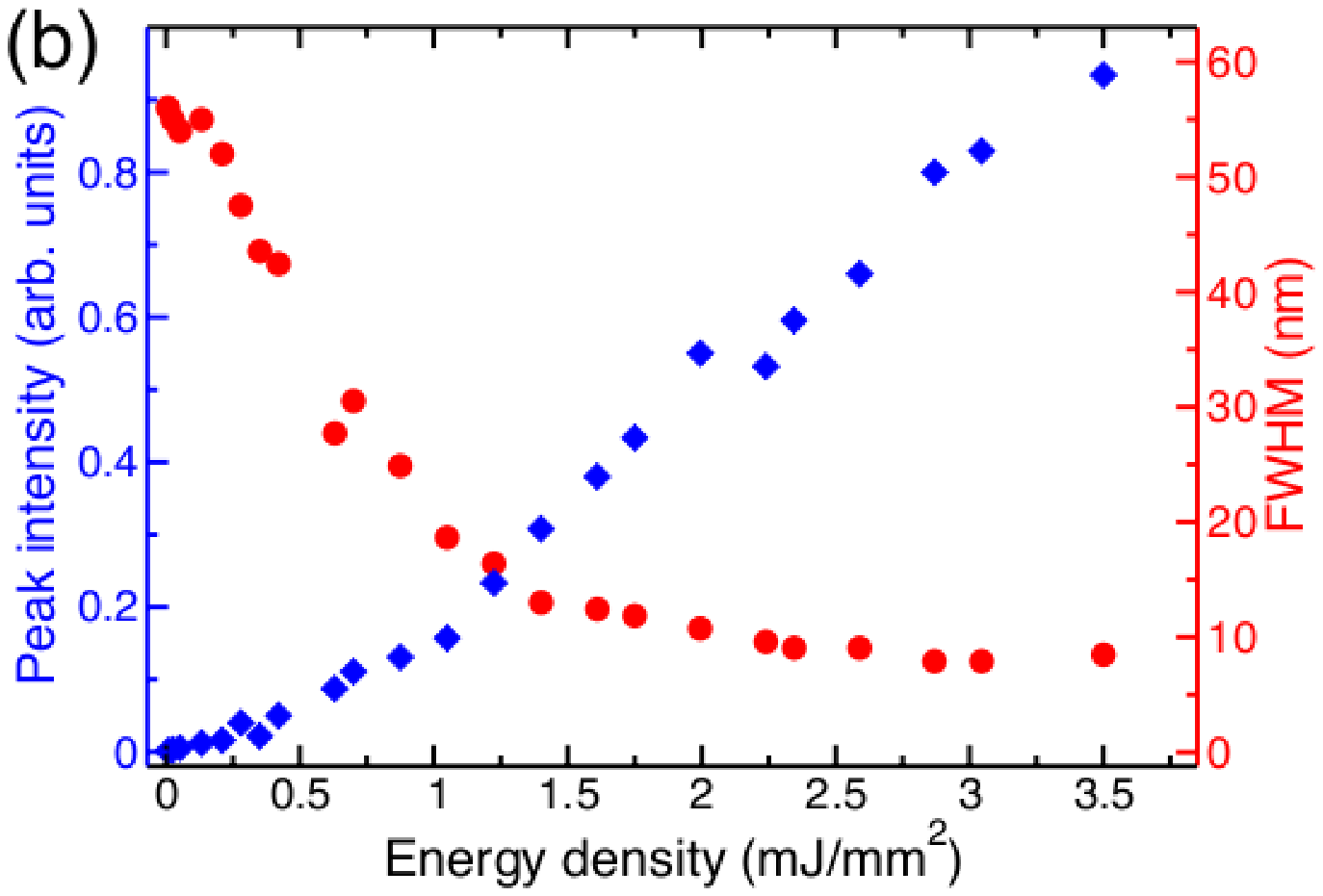}

 \caption{(Color online)  
Emission spectra at different pump energy density for the 300$\mu m$ thick sample (a) and corresponding peak intensity (blue diamonds) and line width (red circles) vs. pump energy (b).} 
\label{fig2}
\end{figure}

 We report in fig.~\ref{fig2}-a the emission spectra from the sample with $L_y=300 ~\mu$m for various injected energy densities (i.e. energy devided to the pump area). The RL coherent emission  is clearly
 shown by the growth of the peak height by two orders of magnitude and by its narrowing for increasing pump, as illustrated in  fig.~\ref{fig2}-b.

\begin{figure}[h]
\includegraphics[width=7cm]{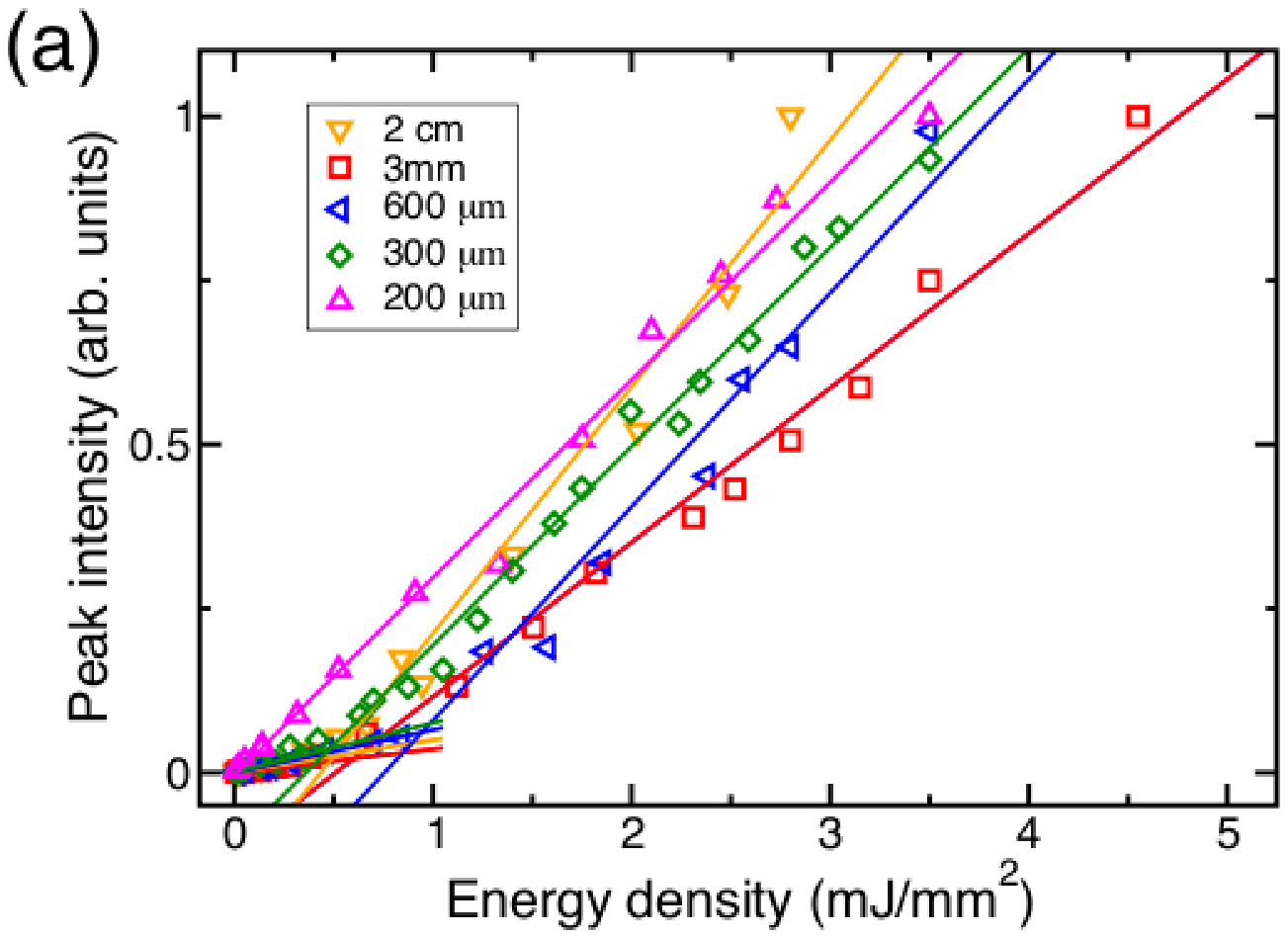}
\includegraphics[width=7cm]{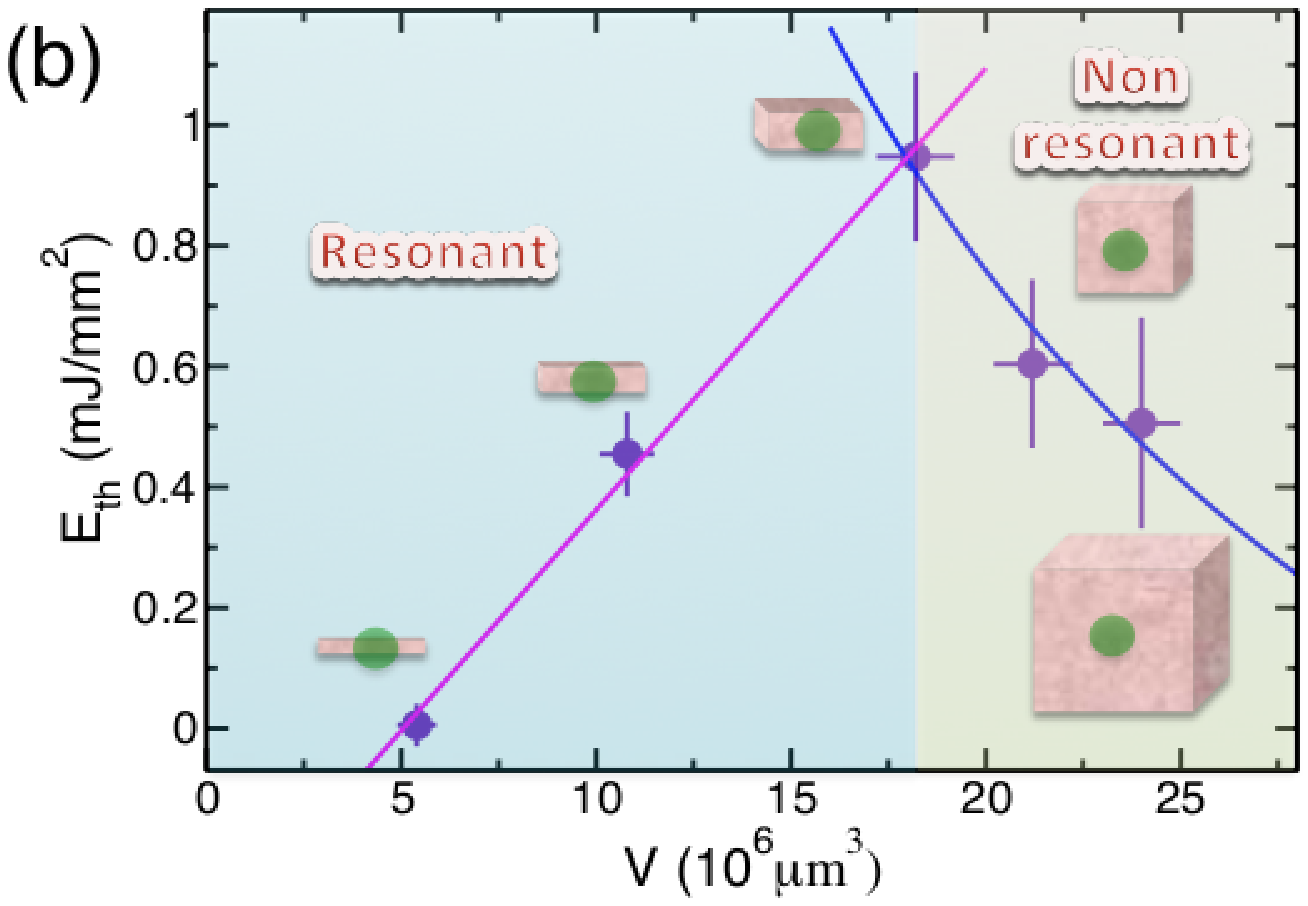}
\caption {(Color online)  
(a) Peaks of the  emitted intensity spectra vs. pump energy density  for samples with different thickness showing different laser thresholds; solid lines are the linear fitting curves.
(b) Energy density threshold $E_{th}$ vs. laser volume $V$   phase diagram showing the resonant - non resonant transition. 
Sketches of the illuminated volumes are reported: the green spot indicates the pumped volume.
Solid lines are the fitting curves after eqs.~(\ref{Ith-NR}) and (\ref{Ith-R}).} 
\label{fig3}
\end{figure}

The single wide emission peak signals the extension of many lasing modes over the pumped volume, as expected for typical weakly scattering RLs.
From the trend of the peak intensity it is clear that there is a  threshold in the pump energy between the slow growth of the spontaneous emission at low energy and
the fast linear increase of the stimulated radiation. 
In fig.~\ref{fig3}-a we illustrate the peak intensity versus input energy  density for the different  fabricated samples with different geometry: the three lithographed channels described above and
two non-lithographed bulk paper pieces with different widths $L_y$ as shown on the figure.
The dependence of the laser threshold is strongly affected by the geometry.
From the intersection of the two linear curves, fitting the data, we estimate the  threshold  energy density $E_{th}$ and in fig.~\ref{fig3}-b
we show it versus 
 the volumes of the different RLs.
 We first consider the case without channels.
We calculate  the volumes of the pumped region for the two bulk paper pieces 
as cylindric volume $V=\pi (L_x/2)^2 \times L_z'$.                            
The data for each sample are given in table~\ref{tab1}
\begingroup
\begin{table*}[h!]

\caption{Sample parameters} \label{tab1}
\begin{center}
\begin{tabular}{cccccc}
 \hline

Type      &  $L_x$ ($\mu$m)          &  $L_y$                                              & $L_z'$ ($\mu$m)         & $V$  (10$^6\mu$m$^3$) & $E_{th}$ (mJ/m$^2$)              \\
\hline
Bulk         & 600$\pm$5                 &   (20.0$\pm$0.5)mm                         &  85$\pm$3                     &   24$\pm$1                               &   0.5$\pm$0.2            \\
Bulk         & 600$\pm$5                 &   (3.0$\pm$0.5)mm                         &  75$\pm$3                    &  21$\pm$1                                &   0.6$\pm$0.1          \\     
Channel  & 600$\pm$5               &   (600$\pm$3)$\mu$m                    &  65$\pm$3                    &  18$\pm$1                                 &   0.9$\pm$0.1             \\
Channel  & 600$\pm$5               &   (300$\pm$3)$\mu$m                    &  60$\pm$3                    &  10.8$\pm$0.7                          &   0.45$\pm$0.07            \\
Channel  & 600$\pm$5               &   (200$\pm$3)$\mu$m                    &  45$\pm$3                    & 5.4$\pm$0.5                             &   0.01$\pm$0.03       \\

\hline
\end{tabular}
\end{center}

\end{table*}
\endgroup

To explain the differences of the volumes, we remark that
for the bulk papers we use different widths to change the effective depth $L_z'$
and thus the  lasing volume: by increasing the width,  the dyed fluid penetrates more deeply along $z$ inside the paper due the capillarity effect.
For the  lithographed channels, we 
 calculate the volume of cylinder with $V=\pi (L_x/2)^2 \times L_z'$ for  $L_y=600\mu$m and of parallelepiped with $V\simeq L_x\times L_y \times L_z'$ for 
$L_y=200 \mu$m and $L_y=300 \mu$m.

In fig.~\ref{fig3}-b two district regimes are evident:  without channels  $E_{th}$ decreases by increasing $V$ as expected for   typical RLs ,  where by increasing the gain volume
light travels a longer path due to more scattering events  and the process of photon generation is amplified. 
Thus, by enlarging the laser volume the losses are more compensated and the threshold is lowered~\cite{vanSoest99}.
This effect has been quantified by Lethokov~\cite{Let68, Cao2003} by considering the diffusive motion of photons in a scattering medium and 
consequently by solving the diffusion equation   for photon energy density $W(r,t)$ in presence of a uniform gain
\begin{equation} \label{eq:diffusion}
\frac{\partial W(r,t)}{\partial t}=D\nabla^2 W(r,t)+\frac{v}{l_g}W(r,t)
\end{equation}
where $D=v l_t/3$ is the diffusion coefficient, $v$ is the photon transport velocity inside the scattering sample and $l_g$ is the gain length (inverse of the gain coefficient $g$).
Equation~(\ref{eq:diffusion}) has the following general solution:
 \begin{equation} \label{sol:diffusion}  
W(r,t)=\sum_n a_n \Psi_n(r)e^{-(D B_n^2-v/l_g)t}
\end{equation}
that passes from an exponential decay to an exponential growth at the threshold condition
\begin{equation} \label{lg}
l_g^{th}=\frac{v}{D B_1^2}=\frac{3}{l_t B_1^2}
\end{equation}
with $B_1$ the lowest of the eigenvalues $B_n$, proportional to the inverse of the average size of the medium $V^{1/3}$.
The proportionality between 
the gain coefficient at threshold $g_{th}=1/l_g^{th}$ and the threshold  energy $E_{th}$ and equation~(\ref{lg}) give the following relation
\begin{equation} \label{Ith-NR}
E_{th}\propto V^{-2/3}.
\end{equation}
In fig.~\ref{fig3}-b the solid line in the right part is the fitting curve to the data, consistent with the experimental results.

Contrarily if the paper is lithographed in channels an opposite behavior is manifested, that is typical of standard lasers with external feedback (mirrors),
 where, even though the  threshold inversion density results $\Delta N_{th}=2 g_{th} /\sigma$, with $\sigma$ the transition cross section and
$ g_{th}=2\delta_c/L_y$, with $\delta_c$ including the contribution of the cavity losses and output coupling, the 
upper-laser-level population  created by the pumping rate is $N_2^{th}\propto P_{th}/V$, proportional to the pump power $P_{th}$ per unit volume,
see~\cite{Siegman}, eq. 10, page 460.
This means that for the same pump power, a smaller volume leads to a larger atomic population inversion resulting to the linear relation   
\begin{equation} \label{Ith-R}
E_{th}\propto  V
\end{equation}
In fig.~\ref{fig3}-b the solid line in the left part is the  linear fitting curve to the experimental points, showing the consistency of the general form~(\ref{Ith-R}) 
 with our data.
 Notice that we use equations~(\ref{Ith-NR}) and~(\ref{Ith-R}) as general descriptions of the two distinct trends we observe in the 
 experimental results and these observed trends can be used for the validation of theoretical models.

In summary in fig.~\ref{fig3}-b  we present a new kind of transition in RLs: from RL where the feedback mechanism is solely given by the scattering effect 
when it is not geometrically  confined and in this sense called 
{\it non-resonant}, to RL  where the same material, constrained in micro-channels with defined walls acting as cavity, shows a laser-like behavior 
because a secondary geometric feedback effect dominates, and for this reason named {\it resonant} RL.
In presence of boundaries  on the transverse
direction, the scattering is strong since the waves cannot
escape from the side of the sample and therefore it is
easy to achieve wave localization. 
 In other words, the  paper channels enhance the scattering effect and
the light has a greater probability of returning to the coherence volume
and interferes with itself and this gives the resonant behavior.
We  stress that the channel walls are present only on the $y$-direction,
while in the $z$-direction the RL (paper+dye) is between air and pure paper.
By investigating the dependence of the pump energy at laser threshold on the laser volume  we depict  the phase diagram of such transition where the two opposite
behaviors of the stimulated emission are clearly shown and quantified by means of general theories of standard and random lasers.
Our results not only evidence the crucial role of geometrical confinement in the physics of disordered lasers, but they also show the potentiality of soft lithographic 
techniques in realizing cheap paper-based photonic devices and open the way to lab on chip and optofluidic integrated systems~\cite{Bha12}.

The research leading to these results has received funding from
 the Italian Ministry of Education, University
and Research under the Basic Research Investigation
Fund (FIRB/2008) program/CINECA grant code
RBFR08M3P4.



\end{document}